\documentclass[a4paper, 12pt]{article}
\usepackage[english]{babel}
\usepackage[a4paper, inner=2cm, outer=2cm, top=3cm, bottom=3cm]{geometry}
\usepackage[dvipsnames]{xcolor}
\usepackage{mathtools}
\usepackage{pstricks}
\usepackage{caption}
\usepackage{graphicx}
\usepackage{amsmath}
\usepackage{amssymb}
\usepackage{mathcomp}
\usepackage{textcomp}
\usepackage{enumitem}
\usepackage{physics}
\usepackage{romannum}
\usepackage[doublespacing]{setspace}
\usepackage{subcaption}
\usepackage{hyperref}

\hypersetup{
  colorlinks   = true, %Colours links instead of ugly boxes
  urlcolor     = Blue, %Colour for external hyperlinks
  linkcolor    = Black , %Colour of internal links
  citecolor    = Black
}

\title{Geodesic congruence in quantum improved spacetimes}
\author{R. Moti and A. Shojai\\
\textit{\small Department of Physics, University of Tehran, Tehran, Iran}}
\date{}

\begin{document}
\maketitle
\pagenumbering{arabic}
\begin{abstract}
We investigate the geodesic deviation equation in the context of quantum improved spacetimes. The improved Raychaudhuri equation is derived, and it is shown that the classical strong energy condition does not necessarily lead to the convergence of geodesics in a congruence in the quantum improved spacetime.
\end{abstract}

\section{Introduction} \label{1}
Weinberg's asymptotic safety conjecture proposed a successful renormalizibility condition for quantum fields at the UV limit which would end to a predictive renormalizable theory of gravity \cite{Weinberg-1st}.
The process of finding a non--Gaussian UV fixed point for the trajectories of essential couplings on the finite dimensional critical surface, which is the ultimate goal of asymptotic safety conjecture, would be simplified by using the functional renormalization group method \cite{Reuter-1st}.
Therefore, instead of functional integration over all spacetime, one can use the effective average action $\Gamma_k$ satisfying the exact renormalization group equation (ERGE),
\begin{equation}
k\partial_k \Gamma_k = \frac{1}{2} \Tr\bigl[ \bigl(\Gamma^{(2)}_k+\mathcal{R}_k\bigr)^{-1} k\partial_k\mathcal{R}_k \bigr] \ .
\end{equation}
This would simplify the way to obtaining the desired quantities. The $\Gamma_k$ is an effective action averaged over the high momentum fluctuation modes. To achieve at this goal, the low momentum fluctuations in the functional integral are suppressed by an arbitrary mass--squared dimension IR--cutoff function $\mathcal{R}_k\propto k^2 R^{(0)}(p^2/k^2)$, where the dimensionless function $R^{(0)}(p^2/k^2)$ is a kind of smeared step function \cite{Reuter & Saueressig}.
Therefore, without any impression by low--momentum fluctuations, the high--momentum fluctuations are integrated out such as a kind of Wilsonian loop.

Truncation of the theory space would decrease the complexity of the infinite dimensional dynamical system of ERGE, and enables one to solve it nonperturbatively although approximately. It has been shown \cite{Bonanno & Reuter 1, Bonanno & Reuter 3} that the Einstein--Hilbert truncation which spanned the theory space on the $\sqrt{g}$ and $\sqrt{g}R$ basis, leads to an anti--screening gravitational running coupling
\begin{equation} \label{Ru. G}
    G(k) = \frac{G(k_{0})}{1 + \omega G(k_{0})(k^{2}-k_{0}^{2})}
\end{equation}
where $\omega = \frac{4}{\pi}(1-\frac{\pi^{2}}{144})$  and the $ k_{0} $ is a reference scale. The Newton's constant  $G_N$ seems to be  a sutiable choice as the reference, hence $ G_{N} \equiv G(k_{0} \to 0)=G_0$ \cite{Bonanno & Reuter 1, Bonanno & Reuter 3} .

Usually people assume the decoupling idea \cite{Reuter & Saueressig, Reuter & Weyer} and improve the coupling constant $G_0$ of the classical theory to the running one. It seems that this could be a proper approximate shortcut from UV to IR limit in a process of searching for the effects of this quantization method.

The cutoff momentum $k$ is a common choice for the scaling parameter of decoupling. On the flat background, this renomalization group parameter can be identified with the inverse of a function of coordinate with dimension of length, $k\propto D^{-1}(x)$. But this coordinate dependent identification is not useful for the curved spacetimes. Introducing a function of all independent curvature invariants as $\chi(\chi_i)$, in which $\chi_{i = 1,\ldots}$ are the independent curvature invariants, seems to be a suitable coordinate independent choice for the scaling parameter in curved spacetimes \cite{Moti & Shojai 3}. Hence, a proper identification in curved spacetime is $ k =\zeta \chi^{-1}(\chi_i)$ where $\zeta$ is some dimensionless constant.
By this coordinate independence property, improving the coupling constant is possible at the level of the action functional as
\begin{equation} 
  S_{I} = \frac{1}{16 \pi} \int \dd[4]{x} \frac{\sqrt{-g}}{G(\chi)}R +\int \dd[4]{x} \sqrt{-g}\mathcal{L}_{m}\ ,
\end{equation}
without any concerns about breaking the general covariance \cite{Moti & Shojai 3}.

As a result, the quantum corrections would emerge dynamically in the gravitational field equation
\begin{equation} \label{IEM}
  G_{\alpha\beta} = 8\pi G(\chi) \tilde{T}_{\alpha\beta} + G(\chi) X_{\alpha\beta}(\chi) \ ,
\end{equation}
where the $ \tilde{T}_{\alpha\beta}=\frac{-2}{\sqrt{-g}}\frac{\delta(\sqrt{-g}\mathcal{L}_m)}{\delta g^{\alpha\beta}} $ is the classical energy--momentum tensor.
 The dynamical effects of the running coupling are contained in
 \[
    X_{\alpha\beta}(\chi) = \Big( \nabla_{\alpha}\nabla_{\beta} - g_{\alpha\beta}\square \Big) G(\chi)^{-1}   - \frac{1}{2} \bigg( R\mathcal{K}(\chi) \frac{\delta\chi}{\delta g^{\alpha\beta}} +
 \]
\begin{equation}
 \partial_{\mu}\Big (R\mathcal{K}(\chi)\frac{\delta\chi}{\delta (\partial_{\mu}g^{\alpha\beta})}\Big) + \partial_{\mu}\partial_{\nu}\Big (R\mathcal{K}(\chi)\frac{\delta\chi}{\delta (\partial_{\nu}\partial_{\mu}g^{\alpha\beta})}\Big ) \bigg ) \ ,
\end{equation}
which explicitly depends on the identification function $\chi$, its derivatives, the running coupling $G(\chi)$ and its derivative through $\mathcal{K}(\chi)\equiv  \frac{2\pdv*{G(\chi)}{\chi}}{G(\chi)^2}$.

It has to be noted that, in general $\chi$ could be a function of all the independent curvature invariants, but an important question is that what is this function? It should be determined via some physical conditions and/or selection rules. This could be energy conditions and the behavior of the geodesic congruence as discussed in this paper, or other physical conditions to be addressed in a forthcoming work.
For simple models like the ones given by $\chi = R$, $\chi = (R_{\alpha\beta}R^{\alpha\beta})^{1/2}$ and $\chi = (R_{\alpha\beta\gamma\sigma}R^{\alpha\beta\gamma\sigma})^{1/2}$ the $X_{\alpha\beta}(\chi)$ term is derived in \cite{Moti & Shojai 3}.
Note that it is quite possible that some curvature invariants like $R$ be identically zero, even though the curvature of the manifold is not zero. For this case, such simple models are ruled out and invariants constructed out of untraced curvature tensors seem to be a better choice (such as the Kretschmann scalar). 
Other choices like invariants obtained from derivatives of curvature tensor (e.g. $R_{\alpha\beta\gamma\delta;\epsilon\sigma}R^{\alpha\beta\gamma\delta;\epsilon\sigma}$) are also possible and maybe useful.

The metric solution of this equation is the one describes the spacetime which fulfills the asymptotic safety conjecture.

The effect of such quantum corrections on the classical solutions are studied for some specific problems like cosmological solutions \cite{Reuter & Weyer, IGR} and black holes and their thermodynamics \cite{Bonanno & Reuter 1, Falls, Moti & Shojai 2}. Here we investigate the effects on the geometrical concepts such as geodesic deviation and the Raychaudhuri equation. To do so, we first evaluate the quantum corrections to the connection in section II. Then, the geodesic deviation and Raychaudhuri equations are derived in the next two sections.

\section{Improved metric connection}
Determining the eligible connection is an unavoidable issue in the process of investigating the behavior of neighboring geodesics. For this purpose, considering $G(\chi) = G_0 / (1+\omega G_0 \zeta^2\chi^2)$,
we expand the improved equation of motion \eqref{IEM} up to first order of quantum corrections
\begin{equation} 
  G_{\alpha\beta}=8\pi G_0 \tilde{T}_{\alpha\beta}+\omega\mathcal{G}_{\alpha\beta}(\chi) +\mathcal{O}(2) \ ,
  \label{app}
\end{equation}
where $\mathcal{G}_{\alpha\beta}(\chi) \equiv \frac{8\pi\zeta ^2}{\chi^2} \tilde{T}_{\alpha\beta}   + \frac{1}{\omega G_0}X_{\alpha\beta}(\chi) $.

Neglecting the higher orders, the term $\mathcal{G}_{\alpha\beta}(\chi)$ behaves as a source of perturbation to the classical (non--improved) Einstein equation. Thus, the approximate solution $g_{\alpha\beta}=\tilde{g}_{\alpha\beta}+q_{\alpha\beta}$ could be a proper solution for this equation where the quantum correction $q_{\alpha\beta}$ generally depends on the chosen symmetries of the spacetime and also on the classical metric  $\tilde{g}_{\alpha\beta}$. 

Using this decomposition into the quantum and classical effects, the Levi--Civita connection $\Gamma^{\gamma}_{\alpha\beta} = \frac{1}{2}g^{\gamma\kappa} (\partial_{\alpha}g_{\kappa\beta}+\partial_{\beta}g_{\alpha\kappa}-\partial_{\kappa}g_{\alpha\beta})$ can be written as
\begin{equation} 
  \Gamma^{\gamma}_{\alpha\beta}=\tilde{\Gamma}^{\gamma}_{\alpha\beta}+ \mathcal{C}^{\gamma \ (QC)}_{\alpha\beta} + \mathcal{C}^{\gamma \ (QQ)}_{\alpha\beta} \ ,
\end{equation}
where the $\tilde{\Gamma}^{\gamma}_{\alpha\beta} = \frac{1}{2}\tilde{g}^{\gamma\kappa} (\partial_{\alpha}\tilde{g}_{\kappa\beta}+\partial_{\beta}\tilde{g}_{\alpha\kappa}-\partial_{\kappa}\tilde{g}_{\alpha\beta})$ is the classical connection. The quantum--classical part
\begin{equation}
  \mathcal{C}^{\gamma \ (QC)}_{\alpha\beta} = -q_{\kappa\sigma}\tilde{g}^{\gamma\sigma}\tilde{\Gamma}^{\kappa}_{\alpha\beta}+\tilde{g}^{\kappa\gamma}\mathcal{Q}_{\alpha\beta\kappa}
\end{equation}
is the contribution of both the classical connection and the quantum part $q_{\kappa\sigma}$ to the full connection.
On the other hand the part $\mathcal{C}^{\gamma \ (QQ)}_{\alpha\beta} \equiv q^{\kappa\gamma}\mathcal{Q}_{\alpha\beta\kappa} $, where $ \mathcal{Q}_{\alpha\beta\kappa} = \frac{1}{2}(\partial_{\alpha}q_{\beta\kappa}+\partial_{\beta}q_{\kappa\alpha}-\partial_{\kappa}q_{\alpha\beta})$, describes the pure quantum corrections and since it is of the second order in  $q_{\alpha\beta}$, we neglect it in what follows.
Also, because $\mathcal{C}^{\gamma \ (QC)}_{\alpha\beta}$ is defined as the difference between two connections, it behaves like a tensor.

\section{Improved geodesic deviation} \label{3}
The behavior of neighboring geodesics is described by the tidal forces which deflect the Euclidean parallel geodesics in curved spacetime. As in general relativity, this behavior can be investigated by considering a 2-surface $\mathcal{S}$ which is covered by a congruence of timelike geodesics. If $\tau$ be an affine parameter along the specified geodesic and distinct geodesics are labeled by a parameter $\lambda$, then $x^{\mu}\equiv x^{\mu}(\tau,\lambda)$ is a parametric equation of the surface $\mathcal{S}$. Any point on this surface is specified by two vector fields, $u^{\mu}= \pdv*{x^{\mu}}{\tau}$ as a tangent vector of the geodesic and $\xi^{\mu}=\pdv*{x^{\mu}}{\lambda}$ which connects two nearby curves. Therefore, on using the Lie derivative relations $\mathcal{L}_{u} \xi^{\mu}=\mathcal{L}_{\xi} u^{\mu}=0$ besides symmetric properties of a connection, one gets
\begin{equation} \label{LieD}
  u^{\beta} \nabla_{\beta}\xi^{\alpha} = \xi^{\beta}\nabla_{\beta}u^{\alpha} \ .
\end{equation}

The geodesic deviation, which describes the relative acceleration of neighboring geodesics is interpreted as the parallel transportation of $\textit{D}\xi^{\alpha}/\textit{D}\tau = u^{\mu}\nabla_{\mu}\xi^{\alpha}$. From eq. \eqref{LieD}, this covariant derivative becomes
\begin{equation} \label{GD1}
  \frac{\textit{D}^2 \xi^{\alpha}}{\textit{D}\tau^2} = -R^{\alpha}_{\ \beta\gamma\sigma} u^{\beta}\xi^{\gamma}u^{\sigma} + \xi^{\beta}\nabla_{\beta} (u^{\gamma}\nabla_{\gamma} u^{\alpha})\ .
\end{equation}
Note that, conservation of stress tensor ($\nabla_\mu\tilde{T}^{\mu\nu}=0$) for a congruence of particles forces the second term to vanish. But 
in the approximation $g_{\mu\nu}=\tilde{g}_{\mu\nu}+q_{\mu\nu}$, the conservation relation reduces to $(\nabla_\mu\tilde{T}^{\mu\nu})_{\textrm{evaluated on } \tilde{g}_{\mu\nu}}+\textrm{other terms linear in } q_{\mu\nu}\simeq 0$. Therefore, particles does not move on the geodesics of $\tilde{g}_{\mu\nu}$, and thus the second term is nonzero up to the first order in $q_{\mu\nu}$. 

On using this approximation and equation \eqref{app}, and
by virtue of Bianchi identity, $\nabla^{\alpha}G_{\alpha\beta}=0$, this term becomes 
\begin{equation}
  u^{\beta}\nabla_{\beta}u_{\alpha} =  -\frac{1}{8\pi} \Bigl(G_{\beta\alpha}\nabla^{\beta} \mathcal{J}(\chi)+\nabla^{\beta} X_{\beta\alpha}(\chi) \Bigr) \ ,
\end{equation}
where $\mathcal{J} \equiv G^{-1}(\chi) $.
Hence, the geodesic deviation equation in the asymptotic safety context is
\begin{equation}
  \frac{\textit{D}^2 \xi^{\alpha}}{\textit{D}\tau^2} = -R^{\alpha}_{\ \beta\gamma\sigma} u^{\beta}\xi^{\gamma}u^{\sigma} - \frac{1}{8\pi}\xi^{\beta}\nabla_{\beta}\nabla^{\gamma} \left[ \mathcal{J(\chi)}g^{\alpha\kappa}G_{\gamma\kappa} +g^{\alpha\kappa} X_{\gamma\kappa}(\chi) \right] \ .
\end{equation}
This shows that, in addition to the tidal forces, the deviation vector would be affected by the variations of the improvement terms. Clearly, this variation depends on the chosen scaling parameter $\chi$. 

The relation can be written more clearly on substituting $g_{\alpha\beta}=\tilde{g}_{\alpha\beta}+q_{\alpha\beta}$, and expanding up to the first order as
\[
\frac{\textit{D}^2 \xi^{\alpha}}{\textit{D}\tau^2} = -\tilde{R}^{\alpha}_{\ \beta\gamma\sigma} \tilde{u}^{\beta}\xi^{\gamma}\tilde{u}^{\sigma}-2\tilde{R}^{\alpha}_{\ \beta\gamma\sigma} \tilde{u}^{\beta}\xi^{\gamma}u^{\sigma \ (QC)}-\Delta^{\alpha}_{\ \beta\gamma\sigma} \tilde{u}^{\beta}\xi^{\gamma}\tilde{u}^{\sigma} +
\]
\begin{equation} \label{geodesic dev.}
  \frac{1}{8\pi}\xi^{\beta}\nabla_{\beta}\nabla^{\gamma} \left[ \mathcal{J(\chi)}g^{\alpha\kappa}G_{\gamma\kappa} +g^{\alpha\kappa} X_{\gamma\kappa}(\chi) \right] \ ,
\end{equation}
where $ \Delta^{\alpha}_{\ \beta\gamma\sigma}\equiv R^{\alpha}_{\ \beta\gamma\sigma}-\tilde{R}^{\alpha}_{\ \beta\gamma\sigma}$ is
\begin{equation}
\Delta^{\alpha}_{\ \beta\gamma\sigma} =  \partial_{\gamma}\mathcal{C}^{\alpha \ (QC)}_{\sigma\beta} - \partial_{\sigma}\mathcal{C}^{\alpha \ (QC)}_{\gamma\beta} + \tilde{\Gamma}^{\alpha}_{\gamma\lambda}\mathcal{C}^{\lambda  \ (QC)}_{\sigma\beta} + \mathcal{C}^{\alpha \ (QC)}_{\gamma\lambda}\tilde{\Gamma}^{\lambda}_{\sigma\beta} - \tilde{\Gamma}^{\alpha}_{\sigma\lambda}\mathcal{C}^{\lambda \ (QC)}_{\gamma\beta} - \mathcal{C}^{\alpha \ (QC)}_{\sigma\lambda}\tilde{\Gamma}^{\lambda }_{\gamma\beta}\ .
\end{equation}
Note that any quantity with a tilde is a classical quantity.

While the term $-\tilde{R}^{\alpha}_{\ \beta\gamma\sigma} \tilde{u}^{\beta}\xi^{\gamma}\tilde{u}^{\sigma}$ in \eqref{geodesic dev.} is the classical one, the others are the effects of the quantum improvement. All the correction terms are proportional to the relative distance and could affect geodesics like the classical case.
The second term on the r.h.s, $-2\tilde{R}^{\alpha}_{\ \beta\gamma\sigma} \tilde{u}^{\beta}\xi^{\gamma}u^{\sigma \ (QC)}$, describes the tidal force between the classical geometry and its quantum fluctuations. The third one, $-\Delta^{\alpha}_{\ \beta\gamma\sigma} \tilde{u}^{\beta}\xi^{\gamma}\tilde{u}^{\sigma}$, determines how two classical parallel geodesics deviate because of the quantum fluctuations of the classical background. It is notable that the $\Delta^{\alpha}_{\ \beta\gamma\sigma}$ is nothing but the derivatives of quantized metric, which is compatible with the geometrical definition of the tidal force. And, the last term at the r.h.s is caused by dynamical self--coupling of geometry and the matter terms.

 This additional quantum effect may be attractive or repulsive depending on the structure and symmetries of the spacetime and the quantum aspects.
This symmetry dependence is the result of the dependence of the cutoff function $\chi(\chi_i)$ on the metric.  It has to be noted that any singular  curvature invariant, should be avoided in $\chi$, at least in the vicinity of the singularity.

This property would be clarified more by studying the evolution of the cross--sectional volume of a congruence of geodesics or Raychaudhuri equation. In the next section we would see that the improved Raychaudhuri equation is so that the strong energy condition could not guarantee the attractive property of quantum improved gravity.

\section{Improved Raychaudhuri equation} \label{4}
The behavior of gravity can be investigated more precisely by studying the evolution of the expansion parameter, which describes the cross--sectional volume of geodesics congruence in the Raychaudhuri equation framework. Although the strong energy condition in the classical Raychaudhuri equation assures the attractive behavior of gravity, we would see that the improvement may change the situation in certain conditions.

As in the classical case, to obtain the Raychaudhuri equation, we first investigate the kinematics of a congruence of geodesics. 

\subsection{Kinematics of a timelike geodesic congruence}
The purely transverse tensor $B_{\alpha\beta}\equiv \nabla_{\beta}u_{\alpha}$ measures the parallel transport failure of $\xi^{\alpha}$ along the congruence. For the linear decomposition \eqref{IEM}, we would have 
\begin{equation}
   B^{\alpha}_{\beta} = \tilde{B}^{\alpha}_{\beta} + \mathcal{B}^{\alpha \ (QC)}_{\beta}
\end{equation}
where the $ \tilde{B}^{\alpha}_{\beta} \equiv \tilde{\nabla}_{\beta}u^{\alpha} $ is the classical form of $B^{\alpha}_{\beta}$ and the $\mathcal{B}^{\alpha \ (QC)}_{\beta} \equiv \mathcal{C}^{\alpha \ (QC)}_{\beta\gamma}u^{\gamma}$ is the quantum correction on it. 

The expansion,
\begin{equation} \label{decomp.}
  B_{\alpha\beta} = \frac{1}{3} \theta h_{\alpha\beta} + \sigma_{\alpha\beta} + \omega_{\alpha\beta}
\end{equation}
where $h_{\alpha\beta}=g_{\alpha\beta}+u_{\alpha}u_{\beta}$ is the spacelike hypersurface induced metric, would facilitate understanding the evolution of deviation vector $\textit{D} \xi^{\alpha}/\textit{D}\tau = B^{\alpha}_{\beta}\xi^{\beta}$.

The scalar expansion $ \theta $ which is the trace of $B_{\alpha\beta}$ becomes
\begin{equation} \label{expansion}
  \theta \equiv B^{\alpha}_{\alpha} = \tilde{\theta} + \vartheta^{\alpha \ (QC)}_{\alpha}
\end{equation}
where for the linear solution $g_{\alpha\beta}=\tilde{g}_{\alpha\beta}+q_{\alpha\beta}$, the term $\vartheta^{\alpha \ (QC)}_{\alpha} = -q^{\alpha\beta}\tilde{B}_{\alpha\beta}+ \mathcal{B}^{\alpha \ (QC)}_{\alpha}$ is the first order correction to the classical expansion parameter $\tilde{\theta} \equiv \tilde{B}^{\alpha}_{\alpha}$.

The shear tensor $\sigma_{\alpha\beta}$ is the traceless symmetric component of the decomposition \eqref{decomp.} and can be expanded as
\begin{equation}
  \sigma_{\alpha\beta} \equiv  B_{(\alpha\beta)}-\frac{1}{3} h_{\alpha\beta}\theta = \tilde{\sigma}_{\alpha\beta} + \varsigma^{\ (QC)}_{\alpha \beta}
\end{equation}
where the $\tilde{\sigma}_{\alpha\beta} $ is the classical shear tensor and the $ \varsigma^{\ (QC)}_{\alpha\beta} = \mathcal{B}^{(QC)}_{(\alpha\beta)} - (\tilde{h}_{\alpha\beta}\vartheta^{\gamma \ (QC)}_{\gamma} + q_{\alpha\beta}\tilde{\theta})/3 $ is the first order quantum correction.

Finally, the last term of $B_{\alpha\beta}$ is the antisymmetric rotation tensor $\omega_{\alpha\beta}$ which up to the first order correction becomes
\begin{equation} \label{rotation}
  \omega_{\alpha\beta} \equiv B_{[\alpha\beta]} = \tilde{\omega} _{\alpha\beta} + w^{\ (QC)}_{\alpha\beta}
\end{equation}
with the correction term $w^{\ (QC)}_{\alpha\beta} = \mathcal{B}^{(QC)}_{[\alpha\beta]}$.

Since $\textit{D} \xi^{\alpha}/\textit{D}\tau = B^{\alpha}_{\beta}\xi^{\beta}$, any small displacement from a spacelike hypersurface to another one would be described by $ \Delta\xi^{\alpha} = B^{\alpha}_{\beta} \xi^{\beta}(t_0) \Delta t$. Three distinct cases can portray each element of this deviation from parallel transportation, best:
\begin{enumerate}[label=(\roman*)]
  \item The dynamical spacetime without any rotation and shearing, or $\sigma_{\alpha\beta} = \omega_{\alpha\beta} = 0$. In this case, only the parameter $\theta$ would be the source of displacement such as
  \begin{equation}
    \Delta \xi^{\alpha} = \frac{1}{3} \theta \xi^{\alpha}(t_0) \Delta t \ ,
  \end{equation}
   which is just the pure expansion. This expansion may results from either the deviation of the classical metric field ($\mathcal{C}^{\alpha \ (QC)}_{\beta\gamma}u^{\gamma}$), or from the interaction of the classical metric $\tilde{g}_{\alpha\beta}$ with the quantum correction $q_{\alpha\beta}$ which has footprints in $\vartheta^{\alpha \ (QC)}_{\alpha}$.
  
It is important to note that the zero value of any $\sigma_{\alpha\beta} $ or $ \omega_{\alpha\beta}$ quantity, does not mean the vanishing of their classical counterparts. Instead, the classical spacetime may have non zero rotating or shearing properties,  canceled by quantum terms.
  \item The dynamical spacetime which does not experience any shearing or expansion, in other words, $\sigma_{\alpha\beta}=0 $ and  $\theta=0$. Hence,
  \begin{equation}
        \Delta \xi^{\alpha} = \omega^{\alpha}_{\beta} \xi^{\beta}(t_0) \Delta t \ .
  \end{equation}
  Since the quantum corrected term $ w^{\ (QC)}_{0i} $ may have nonzero value, this case, unlike the previous one, could end up to asynchronized effects like what happens in Palatini $f(R)$ theories \cite{FShojai & AShojai}.
  
  Same as for the first case, the vanishing values of the elements $\sigma_{\alpha\beta} $ and  $\theta$ does not mean that their classical values are zero.
  \item The last case, is the spacetime which does not make any rotation or expansion between the geodesics, so $\omega_{\alpha\beta}=0$ and $\theta=0$. This assumption leads to
  \begin{equation}
            \Delta \xi^{\alpha} = \sigma^{\alpha}_{\beta} \xi^{\beta}(t_0) \Delta t \ .
  \end{equation}
Again, the quantum correction of transverse vector $\xi^{\alpha}$  to the classical shearing $\tilde{\sigma}_{\alpha\beta}$ is not limited to extra shearing $\varsigma_{ij}$. And asynchronization which is caused by the element  $\varsigma_{0i}$ could be considerable for this kind of spacetimes, too.
\end{enumerate} 

\subsection{Improved Raychaudhuri equation}
To derive the improved Raychaudhuri equation, which describes the evolution of expansion scalar, we begin by studying the time derivation of $B^{\alpha}_{\beta}$. We have
\begin{equation}
  \frac{\textit{D}B_{\alpha\beta}}{\textit{D}\tau} = u^{\gamma}\nabla_{\gamma} B_{\alpha\beta} = u^{\gamma}\tilde{\nabla}_{\gamma}\tilde{B}_{\alpha\beta} + u^{\gamma}\tilde{\nabla}_{\gamma} \mathcal{B}^{(QC)}_{\alpha\beta} - u^{\gamma}\mathcal{C}^{\kappa \ (QC)}_{\gamma\beta}\tilde{B}_{\alpha\kappa} - u^{\gamma}\mathcal{C}^{\kappa \ (QC)}_{\alpha\gamma}\tilde{B}_{\kappa\beta} \ .
\end{equation}
The first correction terms, $u^{\gamma}\tilde{\nabla}_{\gamma} \mathcal{B}^{(QC)}_{\alpha\beta}$, is the contribution of quantum semi--connection evolution. On the other hand, the last two terms describe the interaction of the classical deviation source and the quantum one along the geodesic passage.

Now, the evolution of expansion scalar is obtained by taking the trace of the above equation in the improved spacetime:
\begin{equation}
  \frac{\textit{D}\theta}{\textit{D}\tau} = \frac{\textit{D}\tilde{\theta}}{\textit{D}\tau} - u^{\gamma} \left(q^{\alpha\beta}\tilde{\nabla}_{\gamma}+2\tilde{g}^{\kappa\alpha}\mathcal{C}^{\beta \ (QC)}_{\kappa\gamma}\right)\tilde{B}_{\alpha\beta}+u^{\gamma}\tilde{g}^{\alpha\beta}\tilde{\nabla}_{\gamma}\mathcal{B}^{(QC)}_{\alpha\beta} \ .
\end{equation}
Since
\begin{equation}
  \frac{\textit{D}\tilde{\theta}}{\textit{D}\tau}= - \frac{1}{2} \tilde{\theta}^2 - \tilde{\sigma}^{\alpha\beta}\tilde{\sigma}_{\alpha\beta}+\tilde{\omega}^{\alpha\beta}\tilde{\omega}_{\alpha\beta} - \tilde{R}_{\alpha\beta}u^{\alpha}u^{\beta}
\end{equation}
and
\begin{equation}
  R_{\alpha\beta}=\tilde{R}_{\alpha\beta}+\Delta^{\gamma}_{\alpha\gamma\beta}
\end{equation}
and by considering equations \eqref{expansion}--\eqref{rotation} the improved Raychaudhuri equation is derived as:
\begin{equation} \label{IRE}
   \frac{\textit{D}\theta}{\textit{D}\tau} = \begin{multlined}[t] -\frac{1}{2}\theta^2 - \sigma^{\alpha\beta}\sigma_{\alpha\beta}+\omega^{\alpha\beta}\omega_{\alpha\beta} - R_{\alpha\beta}u^{\alpha}u^{\beta}\\
    +2\theta\vartheta^{\alpha \ (QC)}_{\alpha} +2\sigma^{\alpha\beta}\varsigma^{(QC)}_{\alpha\beta} -2\sigma_{\alpha}^{\ \beta}\sigma_{\gamma\beta}q^{\alpha\gamma}-2\omega^{\alpha\beta}\varsigma^{(QC)}_{\alpha\beta}  + 2\omega_{\alpha}^{\ \beta} \omega_{\gamma\beta}q^{\alpha\gamma} \\
   -\Delta^{\gamma}_{\alpha\gamma\beta}u^{\alpha}u^{\beta} -u^{\gamma}\left[q^{\alpha\beta}\tilde{\nabla}_{\gamma}+2\tilde{g}^{\kappa\alpha}\mathcal{C}^{\beta \ (QC)}_{\kappa\gamma}\right]\tilde{B}_{\alpha\beta} + u^{\gamma}\tilde{g}^{\alpha\beta}\tilde{\nabla}_{\gamma}\mathcal{B}^{(QC)}_{\alpha\beta} \ .
   \end{multlined}
\end{equation}

Although in the improved spacetime the transversity of $\sigma_{\alpha\beta}$ and $B_{\alpha\beta}$ are saved, other terms of this equation can change the standard attractive behavior of gravity and should be investigated separately.

It is interesting to notice that since the first order corrections are considered, all the non--classical terms of the improved Raychaudhuri equation describe the behavior of the quantum fluctuations of the geodesic on the classical background.

Indeed, since the Einstein equation is improved in this context, the strong energy condition does not guarantee the non--negativity of $R_{\alpha\beta}u^{\alpha}u^{\beta}$  for the quantum improved spacetime.

Rewriting the improved field equation \eqref{IEM} as
\begin{equation}
  T_{\alpha\beta}-\frac{1}{2}Tg_{\alpha\beta} = \frac{1}{8\pi} \left(\mathcal{J}(\chi)R_{\alpha\beta} - X_{\alpha\beta}(\chi)+\frac{1}{2} X(\chi)g_{\alpha\beta} \right) \ ,
\end{equation}
the strong energy condition $(T_{\alpha\beta}-\frac{1}{2} T g_{\alpha\beta})u^{\alpha}u^{\beta} \geqslant 0$ results in 
\begin{equation}
  R_{\alpha\beta}u^{\alpha}u^{\beta} \geqslant \frac{1}{\mathcal{J}(\chi)} \left( X_{\alpha\beta}(\chi)u^{\alpha}u^{\beta}-\frac{1}{2}X(\chi) \right)
\end{equation}
where $X(\chi)$ is the trace of $X_{\alpha\beta}(\chi)$. Therefore, to get the condition $R_{\alpha\beta}u^{\alpha}u^{\beta} \geqslant 0$, the non--negativity of the quantum term $ (X_{\alpha\beta}(\chi)u^{\alpha}u^{\beta}-\frac{1}{2}X(\chi))/\mathcal{J}(\chi)$, is necessary. This is not what would happen always for any spacetime and any choice of $\chi$.

\section{Concluding remarks}\label{6}
The asymptotic safety conjecture predicts a renormalizable field theory when there is a non--Gaussian fixed point at the UV limit of  the theory. By using the functional renormalization group methods, with some consideration, one would end to an antiscreening running gravitational coupling which has a non--Gaussian fixed point needed by this conjecture.
It is shown that a general function of curvature invariants, $\chi(\chi_i)$, seems to be a suitable scaling parameter for gravity theory as a renormalizable quantum field theory \cite{Moti & Shojai 3}. Considering the running gravitational coupling $G(\chi)$, and improving the Einstein--Hilbert action would result in the improved equation of motions which contain this kind of quantum correction effects.

In this paper, we investigated the properties of a geodesics congruence in this context. First the geodesic deviation equation is derived. It is shown that the deviation contains improvement terms.
The improved Raychaudhuri equation shows that the strong energy condition, unlike the classical general relativity, does not necessarily end to a converging quantum corrected geodesic congruence, because of the presence of correction terms.

To make this more clear, consider a congruence of timelike geodesics which are hypersurface orthogonal. Since the improved equation of motion \eqref{IEM} results in
\begin{equation}
  R_{\alpha\beta}=\frac{8\pi}{\mathcal{J}(\chi)} (T_{\alpha\beta}-\frac{1}{2}Tg_{\alpha\beta}) + \frac{1}{\mathcal{J}(\chi)} (X_{\alpha\beta}-\frac{1}{2}Xg_{\alpha\beta}) \ ,
\end{equation}
the convergence of this congruence depends on the condition 
\begin{equation} \label{CGC}
 \begin{multlined}[t] \frac{8\pi}{\mathcal{J}(\chi)} (T_{\alpha\beta}-\frac{1}{2}Tg_{\alpha\beta})u^{\alpha}u^{\beta} + \frac{1}{\mathcal{J}(\chi)} (X_{\alpha\beta}-\frac{1}{2}Xg_{\alpha\beta})u^{\alpha}u^{\beta} \\
    +2\theta\vartheta^{\alpha \ (QC)}_{\alpha} +2\sigma^{\alpha\beta}\varsigma^{(QC)}_{\alpha\beta} -2\sigma_{\alpha}^{\ \beta}\sigma_{\gamma\beta}q^{\alpha\gamma}-2\omega^{\alpha\beta}\varsigma^{(QC)}_{\alpha\beta}  + 2\omega_{\alpha}^{\ \beta} \omega_{\gamma\beta}q^{\alpha\gamma} \\
   -\Delta^{\gamma}_{\alpha\gamma\beta}u^{\alpha}u^{\beta} -u^{\gamma}\left[q^{\alpha\beta}\tilde{\nabla}_{\gamma}+2\tilde{g}^{\kappa\alpha}\mathcal{C}^{\beta \ (QC)}_{\kappa\gamma}\right]\tilde{B}_{\alpha\beta} + u^{\gamma}\tilde{g}^{\alpha\beta}\tilde{\nabla}_{\gamma}\mathcal{B}^{(QC)}_{\alpha\beta} \geqslant 0  \ .
   \end{multlined}
\end{equation}

As an example, the $X_{\alpha\beta}(\chi)$ for a special case $\chi=R_{\alpha\beta}R^{\alpha\beta}$ would be \cite{Moti & Shojai 3} 
\begin{multline}
X_{\alpha\beta}(\chi) = 
\bigl(\nabla_{\alpha}\nabla_{\beta} - g_{\alpha\beta} \square\bigr) \mathcal{J}(\chi)
+ \mathcal{K}(\chi) R R_{\alpha\kappa} R_{\beta\gamma} g^{\kappa\gamma} \\   
- \nabla^{\gamma}\nabla_{\beta}( \mathcal{K}(\chi) R R_{\alpha\gamma})
 +\frac{1}{2}\square( \mathcal{K}(\chi) R R_{\alpha\beta}) + \frac{1}{2} g_{\alpha\beta} \nabla_{\gamma}\nabla_{\kappa} ( \mathcal{K}(\chi) R R^{\gamma\kappa})
\end{multline}
where $\mathcal{K}(\chi)\equiv  \frac{2\pdv*{G(\chi)}{\chi}}{G(\chi)^2}$.
Up to the first order, only the first term should be considered. Hence, $X=-\square \mathcal{J}(\chi)$ and equation \eqref{CGC} reduces to
\begin{multline}
 \frac{8\pi}{\mathcal{J}(\chi)} (T_{\alpha\beta}-\frac{1}{2}Tg_{\alpha\beta})u^{\alpha}u^{\beta}
  +\frac{1}{\mathcal{J}(\chi)}(\nabla_{\alpha}\nabla_{\beta}-\frac{1}{2}g_{\alpha\beta}\square )\mathcal{J}(\chi)u^{\alpha}u^{\beta} \\
    +2\theta\vartheta^{\alpha \ (QC)}_{\alpha} +2\sigma^{\alpha\beta}\varsigma^{(QC)}_{\alpha\beta} -2\sigma_{\alpha}^{\ \beta}\sigma_{\gamma\beta}q^{\alpha\gamma}-2\omega^{\alpha\beta}\varsigma^{(QC)}_{\alpha\beta}  + 2\omega_{\alpha}^{\ \beta} \omega_{\gamma\beta}q^{\alpha\gamma} \\
   -\Delta^{\gamma}_{\alpha\gamma\beta}u^{\alpha}u^{\beta} -u^{\gamma}\left[q^{\alpha\beta}\tilde{\nabla}_{\gamma}+2\tilde{g}^{\kappa\alpha}\mathcal{C}^{\beta \ (QC)}_{\kappa\gamma}\right]\tilde{B}_{\alpha\beta} + u^{\gamma}\tilde{g}^{\alpha\beta}\tilde{\nabla}_{\gamma}\mathcal{B}^{(QC)}_{\alpha\beta} \geqslant 0  \ .
   \end{multline}
This shows that it is quite possible that for a general improved spacetime, the correspondence of attractiveness and strong energy condition, breaks down.

\vglue1cm
\textbf{Acknowledgment:} This work is supported by a grant from Iran National Science Foundation (INSF).

\end{document}